\documentclass[fleqn,usenatbib]{mnras}

\usepackage{newtxtext,newtxmath}
\usepackage[T1]{fontenc}

\usepackage{graphicx}	% Including figure files
\usepackage{amsmath}	% Advanced maths commands
\usepackage{upgreek}    % Roman Greek symbols (mu for micrometers)
\usepackage{xcolor}     % For highlighting during preparation and review
\title[All-fibre wavefront sensor]{All-fibre wavefront sensor}

\author[T.~A.~Wright et al.]{
T.~A.~Wright,$^{1}$\thanks{E-mail: t.wright@bath.ac.uk}
S.~Yerolatsitis,$^{1,2}$
K.~Harrington,$^{1}$
R.~J.~Harris,$^{3}$
and T.~A.~Birks$^{1}$\thanks{\,E-mail: t.a.birks@bath.ac.uk}
\\
$^{1}$Centre for Photonics and Photonic Materials, Department of Physics, University of Bath, Bath, BA2 7AY, UK\\
$^{2}$Now at CREOL, The College of Optics and Photonics, University of Central Florida, Orlando, Florida 32816, USA\\
$^{3}$Max-Planck-Institute for Astronomy, K\"onigstuhl 17, 69117, Heidelberg, Germany
}

\date{Accepted XXX. Received YYY; in original form ZZZ}

\pubyear{2022}

\begin{document}
\label{firstpage}
\pagerange{\pageref{firstpage}--\pageref{lastpage}}
\maketitle

\begin{abstract}
We report on a tapered three-core optical fibre that can be used as a tip-tilt wavefront sensor. In this device, a coupled region of a few millimetres at the sensing tip of the fibre converts fragile phase information from an incoming wavefront into robust intensity information within each of the cores. The intensity information can be easily converted to linear wavefront error over small ranges, making it ideal for closed loop systems.The sensor uses minimal information to infer tip-tilt and is compatible with remote detector arrays. We explore its application within adaptive optics and present a validation case to show its applicability to astronomy.
\end{abstract}


%%%%%%%%%%%%%%%%%%%%%%%%%%%%%%%%%%%%%%%%%%%%%%%%%%

%%%%%%%%%%%%%%%%% BODY OF PAPER %%%%%%%%%%%%%%%%%%

%%%%%%%%%%%%%%%%%& INTRODUCTION %%%%%%%%%%%%%%%%%%
\section{Introduction}

To overcome the limits imposed by the atmosphere, it is common for ground-based telescopes use adaptive optics (AO) to correct the effect of atmospheric turbulence on incoming light from astronomical objects. This implementation of AO systems has led to numerous previously unachievable scientific discoveries, including the determination of stellar orbits around in the centre of the Milky Way \citep{ghez_1998} and direct imaging detection of planetary systems around other stars \citep{chauvin_2004}. AO systems also offer significant advantages to applications outside of astronomy. AO can correct for atmospheric aberrations in other applications, such as in free-space optical communications \citep{weyrauch_2005,gregory_2012}, or even produce ultrafast laser filamentation for remote location sensing \citep{daigle_2009}. AO has also been used to mitigate aberrations caused by the refractive index structure of samples in a range of biomedical applications such as ophthalmology \citep{akyol_2021}, in-vivo imaging \citep{zawadzki_2005} and microscopy \citep{booth_2014}.

For the AO system to construct the optimal image in the focal plane, two key components are required: a method for sensing the incoming wavefront, and then a method for correcting it. Corrections to the incident wavefronts are commonly enacted by AO systems using one or more deformable mirrors (DMs). For large telescope installations with strict wavefront correction requirements, a state-of-the-art DM may be driven by thousands of actuators, each able to apply a small local perturbation to the mirror on millisecond time scales \citep[table 4]{guyon_2018}. 

The specific instructions for these actuators must be rapidly generated in response to the incoming wavefront's existing state as measured by the wavefront sensor (WFS). Accordingly, the removal of wavefront distortions is limited by how well the WFS describes the current state of the wavefront and how quickly the DM receives and acts on this information. It is common practice to position the WFS in the pupil plane by reimaging the incident wavefront via a dichroic beam splitter. One problem with this approach is that it can introduce path-difference aberrations caused by non-common optical components in the wavefront-sensing and science beams \citep{sauvage_2007}. However, for convenience most WFSs in use do sense the wavefront in the pupil plane, including the Shack–Hartmann (SH) WFS \citep{platt_2001}, the pyramid WFS \citep{ragazzoni_1996}, and the curvature WFS \citep{roddier_1988}. Phase-diversity WFSs, while uncommon in astronomy, are an exception to this trend and operate in the focal plane, thus avoiding path-difference errors \citep{gonsalves_1979,paxman_2007}. We refer the interested reader to the reviews of AO systems given by \citet{davies_2012} and \citet{rigaut_2015}.

Recent studies have detailed attempts to exploit advances in fibre-optic and waveguide technologies to introduce alternative WFS implementations \citep{corrigan_2016,corrigan_2018,norris_2020}. These proposed techniques make use of photonic lanterns \citep{birks_2015} to acquire both phase and amplitude information directly from the focal plane, although the recovery of the wavefront is non-trivial. \citeauthor{norris_2020}, for example, use a deep neural network to reconstruct information on the input phase from the output intensities, such that it is difficult to recover the performance metrics for WFS. More straightforward approaches to recovering tip-tilt information using a photonic lantern are possible \citep{cruz_2021}. However, in such cases, the complex transfer matrix for each lantern still needs to be solved numerically. Furthermore, photonic lanterns are not wavelength agnostic and will likely prove challenging to scale up to measure complex wavefront distortions and to have larger detection windows. This requirement will necessitate using higher mode numbers, making each device harder to calibrate, and using multiple lanterns that would each require individual calibration. Nevertheless, fibre- and waveguide-based techniques provide further advantages, such as allowing users to remap the output of the WFS into any shape, including 1:1 mode-to-pixel configurations. In contrast to the bulky two-dimensional pixel arrays currently used for Shack–Hartmann WFSs (SH-WFSs), these devices are compatible with linear detector arrays located far from the pupil or focal plane.

We report an easily-scalable all-fibre tip-tilt sensor design based on a locally tapered three-core fibre and demonstrate operation in the pupil plane. At the tapered tip, our sensor uses minimal information to infer tip-tilt and converts phase information to amplitude information. This information is then guided through the uncoupled three-core fibre, making the sensor compatible with remote detector arrays. We present experimental verification of the photonics by detecting a diverging beam at a large distance, approximating a flat wavefront. We make these measurements without focusing optics ahead of the fibre probe to make the experimental results more general and not specific to a particular telescope arrangement. Later, we discuss the WFS's response to light in the context of real-world telescope integration and, particularly, how the angular sensitivity is relevant to astronomy. Furthermore, we indicate how to upscale our WFS design to an array of tip-tilt sensors that can deduce higher-order wavefront aberrations.

\section{Principle of Operation}
In an optical system, phase measurements at any three non-collinear points are sufficient to determine the wavefront tip and tilt. In principle, we can use a three-core fibre to transport these phases to a remote location, where they can be analysed, e.g., interferometrically. In practice, the phase information will be lost due to time-dependent and effectively-random inhomogeneities, bends and environmental disturbances to the optical fibre. However, we can exploit core-to-core coupling in a short input section to convert the phase information into amplitude information, which is robustly preserved along the arbitrarily-disturbed fibre length if the fibre is designed so that the cores are uncoupled everywhere except at the input.

\begin{figure}
	\includegraphics[width=\columnwidth]{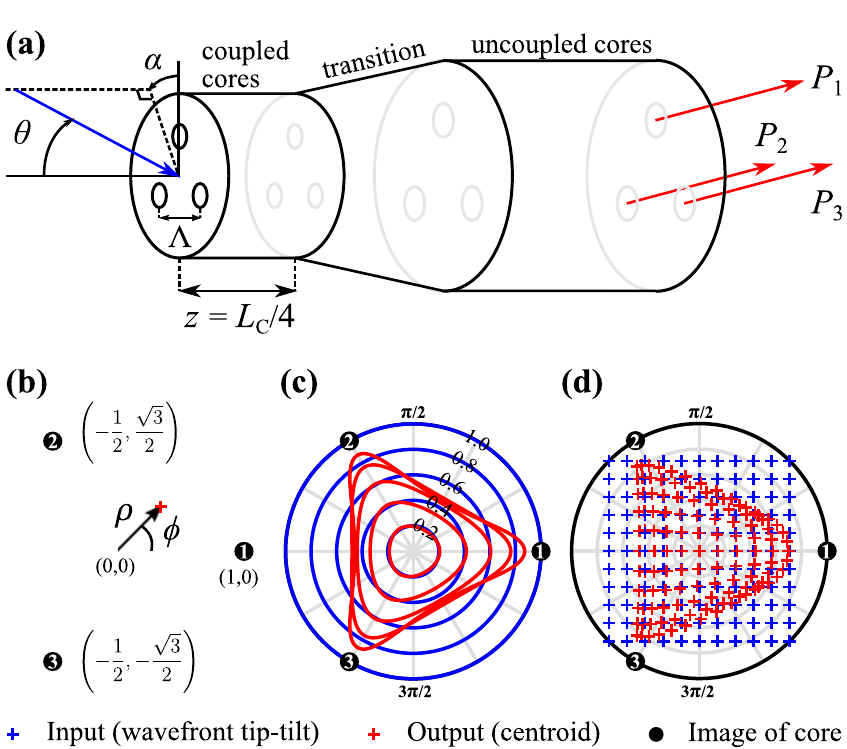}
    \caption{(a) The all-fibre wavefront sensor. The coupled and transition sections are about a centimetre long, while the uniform fibre with uncoupled cores can be several metres long. (b) The centroid of the output powers (red cross) with polar coordinates $(\rho,\phi)$ in a space spanned by normalised core positions, which matches the weighted centroid of a near-field image at the fibre output. (c) Simulated polar plots of output centroid $(\rho,\phi)$, in red, for conically precessing inputs $(\delta,\alpha)$, in blue. (d) Simulated polar plots of output centroid $(\rho,\phi)$, in red, for a square array of input tilts $(\delta,\alpha)$, in blue. We assume $z/L_{\mathrm{C}} = 1/4$.}
    \label{fig:1}
\end{figure}

Depicted in  Fig.\,\ref{fig:1}\,(a), our design consists of a fibre with three identical cores at the corners of an equilateral triangle. We taper (narrow down a short length of) the fibre to locally induce coupling, then cleave it to provide an input end-face: the sensing tip. The spacing of the three cores is otherwise sufficiently large that no coupling occurs between them along the fibre beyond this short tapered region. At the output, we can then monitor the powers in the cores using an image sensor or, in principle, three photodiodes. Figure~\ref{fig:1}\,(a) also shows parameters describing the tilt of the wavefront incident on the fibre end-face: the angle of tilt is $\theta$, and the azimuth of the plane of incidence is $\alpha$. The device’s response, manifested as different powers at each core’s output, can be represented by an intensity-weighted ‘centroid’ position 
\begin{equation}
\mathbf{r}=\frac{P_{1} \mathbf{r}_{1}+P_{2} \mathbf{r}_{2}+P_{3} \mathbf{r}_{3}}{P_{1}+P_{2}+P_{3}}.
\label{eq.1}
\end{equation}
This abstract space is spanned by core position vectors $\mathbf{r}_{\mathrm{n}}$ on a unit circle, as shown in Fig.\,\ref{fig:1}\,(b) where $\mathbf{r}$ is expressed in polar coordinates $(\rho,\phi)$. The full analytical relation between the parameters sought $(\theta,\alpha)$ and measured $(\rho,\phi)$ is described in the Appendix. For sufficiently small tilt angles $\theta$ it can be shown that 
\begin{align}
\rho = \delta, && \phi = \alpha,
\label{eq.2}
\end{align}
and thus, the full expression reduces to
\begin{equation}
\delta=\left[\frac{2 \pi}{\sqrt{3}} \frac{\Lambda}{\lambda} \sin \left(\frac{2 \pi z}{L_{\mathrm{C}}}\right)\right] \theta.
\label{eq.3}
\end{equation}
Here $\delta$ is the tilt angle $\theta$ scaled by a function of the tapered fibre's core separation $\Lambda$, length $z$ and coupling length $L_{\mathrm{C}}$, along with the free-space wavelength $\lambda$ of the light. That is, to a first-order approximation, the displacement, $\rho$, of the centroid is proportional to the input tilt, $\theta$, and the azimuths of the centroid and the plane of incidence are the same. The sensitivity of $\rho$ to $\theta$ is maximised when the length is 1/4 of the coupling length: $z/L_{\mathrm{C}} = 1/4$. From here on, we will assume this to be the case. 

To illustrate, Fig.\,\ref{fig:1}\,(c) depicts the loci of a conically precessing input (i.e. fixed $\delta$ and varying $\alpha$) in blue and the resulting output centroid in red. Five input values of $\delta$, labelled, are plotted. Figure ~\ref{fig:1}\,(d) shows how a square grid of input states maps onto a grid of output centroids. As seen in Fig.\,\ref{fig:1}\,(c) and Fig.\,\ref{fig:1}\,(d), the first-order approximation starts to fail around $\delta = 0.2$, but the relationship remains analytic. Beyond $\delta = 0.8$, we find that the measured centroid folds in on itself, and by $\delta = 1.6$, the sign of the centroid switches. Therefore, we have three correction regimes, the first of which is linear $(\delta < 0.2)$. The second regime $(0.2 <\delta<0.8)$ is no longer linear, but the correction is analytically derivable and the magnitude is correct. The third regime $(0.8<\delta<1.6)$ is also analytical, and the sign of correction is correct, but the magnitude is not. Beyond these three regimes, when $\delta>1.6$, the correction sign changes making the system unstable. Such limits are not unique to our device; this effect is similar to spots wandering between subapertures in a Shack-Hartmann, or non-illumination of one or more quadrants on a pyramid WFS.  

The operation of our WFS is notable for using the least information possible (three powers) to infer tip-tilt. Furthermore, the mapping between the input wavefront perturbation and the output parameters is computationally straightforward. The simplicity of this mapping is in contrast to other fibre wavefront sensors \citep{corrigan_2016,corrigan_2018,norris_2020}, and would be advantageous when attempting to mitigate latency within an integrated astronomical AO system.

\section{Fabrication of the Fibre and WFS}
We fabricated the three-core fibre using the stack-and-draw technique \citep{russell_2003}. Each core has a diameter of 9.5\,$\upmu$m and an NA of 0.11 (single-mode for $\lambda > 1.37\,\upmu$m) in the final fibre. For all of our optical measurements, we used a light source with a wavelength of 1.55\,$\upmu$m. The separation between each core is 38\,$\upmu$m, which is sufficient for us to observe no coupling between cores along a 100-m length of the fibre. This length exceeds any requirement for remote detection in astronomy as it is also vital to minimise latency in the data transmission from a WFS to the deformable mirror.  An optical micrograph of the fibre cross-section is inset within Fig. 3(b).

\begin{figure}
	\includegraphics[width=\columnwidth]{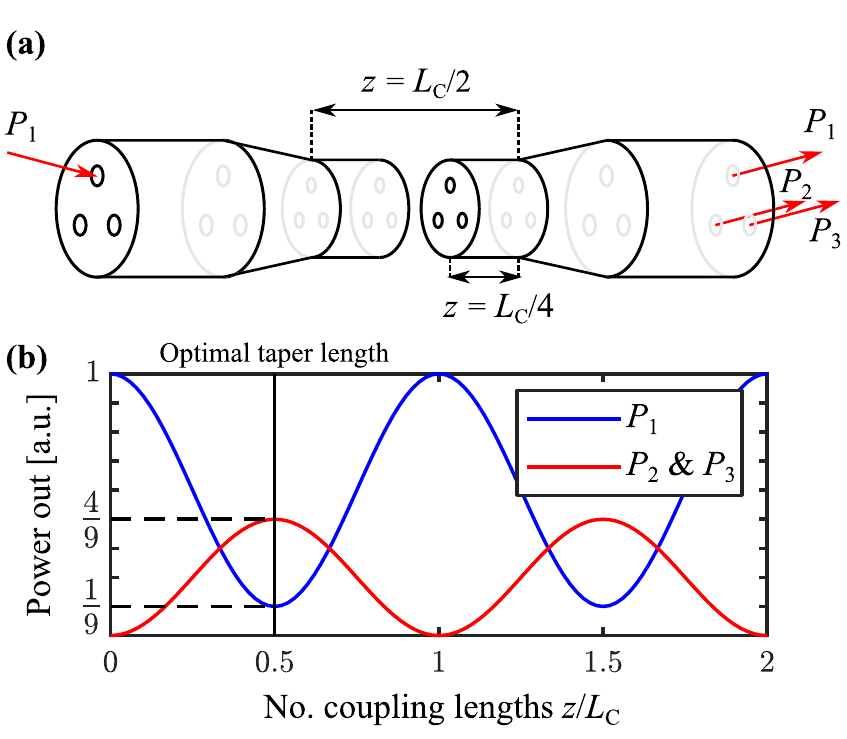}
    \caption{(a) Schematic of the WFS fabrication principle. (b) Simulated variation with length of the output powers in the cores, for input in core 1 only. We taper the three-core fibre whilst monitoring the output using a camera. We couple light into a single core at the input and then taper the fibre until the power in this core at the output is 1/9 of the total, and the power in each of the two remaining cores is 4/9 of the total. At this point, indicated "Optimal taper length" in (b), the symmetric taper has a length of $L_{\mathrm{C}}/2$. We then cleave halfway along the length to form a sensing tip with a length of $L_{\mathrm{C}}/4$. }
    \label{fig:2}
\end{figure}

To form the tapered section, we heat and stretch the fibre, cleaving it such that the length of the coupled region, $z$, is $L_{\mathrm{C}}/4$. To do this, we couple light into a single core and monitor the output of the fibre whilst it is being tapered. As shown in Fig.\,\ref{fig:2}\,(b), the power distribution between cores at the output changes with taper elongation. By stopping the tapering process when the power in core 1 is minimised (1/9 of the maximum value), we ensure that the tapered section's total length is $L_{\mathrm{C}}/2$. Cleaving the tapered fibre at its midpoint, Fig.\,\ref{fig:2}\,(a), produces two equal sections of length $L_{\mathrm{C}}/4$. If we were to cleave at a point offset from the centre, the WFS would still work, although would operate but with a reduced sensitivity to wavefront tilt as described by Eq.~(\ref{eq.3}) when $z \neq L_{\mathrm{C}}/4$.

For the device presented, the combined length of the coupled region and transition was 12\,mm, and the remaining length of the fibre, where the cores were uncoupled, was 1.2\,m. The core-to-core separation in the tip, $\Lambda$, was 24\,$\upmu$m.

\section{Experimental Verification}
\begin{figure}
	\includegraphics[width=\columnwidth]{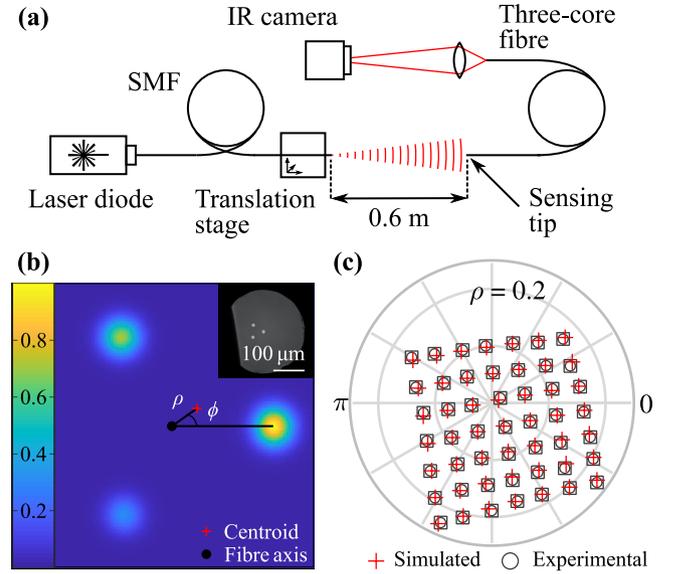}
    \caption{(a) Schematic showing the experimental setup. (b) A typical near-field image used to calculate the centroid $(\rho,\phi)$ of the powers in each core. Inset: Optical micrograph of the untapered three-core fibre. (c) Polar $(\rho,\phi)$ plot of experimental and simulated response of centroid position to input tilts on a square array with pitch 2.9 arcminutes; the plotted range $(\rho < 0.25)$ is roughly equivalent to the central 9 points in Fig.\,\ref{fig:1}\,(d).}
    \label{fig:3}
\end{figure}

To experimentally validate our design, we launched 1.55-$\upmu$m laser light into free space from a single-mode fibre (SMF) mounted on an x-y translation stage positioned 0.6 m from the sensing tip. At this distance, compared with the 24-$\upmu$m separation of the fibre cores, the light emitted from the launch fibre approximated a plane wave at the input end face. Therefore, by adjusting the stage's transverse position, we could control the tip and tilt of this incoming plane wave relative to the device (see in Fig.\,\ref{fig:3}\,(a)). 
To evaluate the centroid positions $(\rho,\phi)$, we collected near-field images of the fibre output using a ×10 lens and a short-wave infrared camera (Raptor Photonics, OWL\,640\,M). This was for convenience: to use a three-pixel linear detector array, we would first need to reformat the fibre output. An example image is shown in Fig.\,\ref{fig:3}\,(b). We adjusted the x-y translation stage to 49 positions within a $7\!\times\!7$ grid, producing step sizes in $\theta$ equal to 2.9 arcminutes, and captured output images at each point. The grid alignment to the fibre was at an arbitrary offset in $\theta$ and $\alpha$. We evaluated the centroid positions from the acquired images and plotted these in Fig.\,\ref{fig:3}\,(c), along with analytical values of $\rho$ and $\phi$. We calculated the centroid positions $(\rho,\phi)$ of the analytical grid using equations 2 and 3, acting on a perfectly square input array $(\delta,\alpha)$. We determined the centre offset and orientation of this square input grid relative to the fibre end-face by taking the mean position of the experimental values of $\rho$ and $\phi$, as well as the gradients of lines drawn through each row and column. In calculating the $\delta$ component of the analytical centroid coordinates, we used the measured values of $\Lambda$ (24 $\upmu$m) and $\lambda$ (1.55 $\upmu$m), and assumed the sinusoidal term was equal to 1 (i.e. $z = L_{\mathrm{C}} /4$). 
\begin{figure}
	\includegraphics[width=\columnwidth]{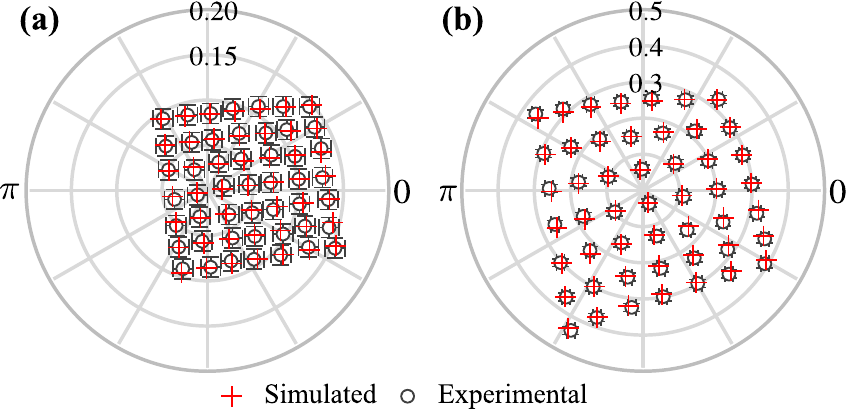}
    \caption{Experimental and simulated response of centroid position to input tilts on a square array with pitch (a) 1.7 arcminutes, and (b) 5.7 arcminutes.}
    \label{fig:4}
\end{figure}
Further to this, we repeated the measurement to record our devices' response to input tilts on grids with tilt spacings of both 1.7 and 5.7 arcminutes, shown in Fig.\,\ref{fig:4}\,(a) and Fig.\,\ref{fig:4}\,(b), respectively.  In all cases, we see good agreement between our experimental and analytical results so that the wavefront tip-tilt can be inferred from the measured centroid positions. The plotted uncertainties relate to the mechanical precision of the x-y stage that we used to control the relative tip-tilt of the plane wave. We measured the change in centroid positions after repeated adjustments of this transverse stage from, and back to, an arbitrarily selected position and found that the recorded centroids were distributed around this point with a standard deviation in $\rho$ of $\sigma$ = 0.0055, which corresponds to $\sim20$\,arcseconds. In Fig.\,\ref{fig:3}\,(c), Fig.\,\ref{fig:4}\,(a) and Fig.\,\ref{fig:4}\,(b), the error bars correspond to $2\sigma = 0.011$.

We have shown that, in our verification setup, our WFS can resolve input tilt separations of $\leq$ 1.7 arcminutes. 
However, in principle, the tilt resolution of our WFS is limited only by the detector's capability to determine the power difference between each core. 
This characteristic demonstrates an optical improvement over a SH-WFS, where the angular resolution at the focal plane is limited by the pixel period and microlens' focal length \citep{Valente_15}. 
For example, the Thorlabs WFS40-5C is a commercial WFS with an effective focal length of 4.1 mm and a pixel pitch of 5.5\,$\upmu$m, thus giving an angular resolution of $\delta\theta = 2.56$ arcminutes. 
Similarly, the SH-WFS on the CANARY AO testbed \citep{gendron_2011} uses a camera with 24-$\upmu$m pixels and microlenses with a focal length of 29 mm; this amounts to an angular resolution of $\delta\theta \sim 2.84$ arcminutes, although higher resolutions can be obtained at a sub-pixel level \citep{tian_1986}.

\section{Use with Astronomical Systems}
\subsection{Telescope integration}
\begin{figure}
	\includegraphics[width=\columnwidth]{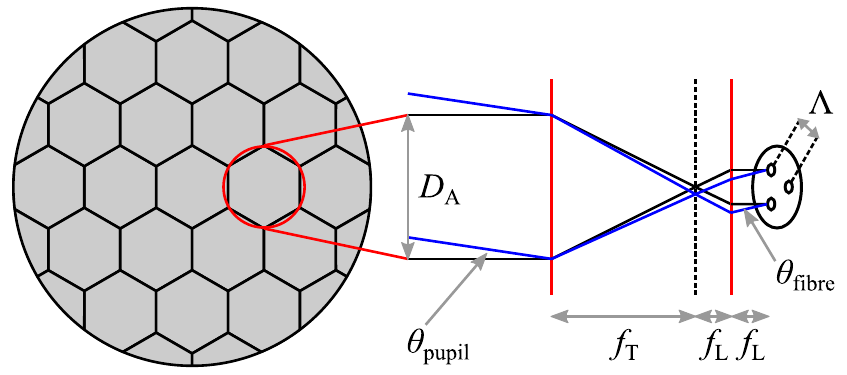}
    \caption{Our model wavefront sensing system arranged in a 4f configuration. On the left a pupil is split into subapertures $D_{\mathrm{A}}$ (here hexagonal segments, though this is frequently square). This is imaged by the telescope of focal length $f_{\mathrm{T}}$, and then the pupil is re-imaged using a microlens of focal length $f_{\mathrm{L}}$. A tilted beam is shown as green dashed lines, with angles at the pupil, $\theta_{\mathrm{pupil}}$, and the fibre, $\theta_{\mathrm{fibre}}$, shown.}
    \label{fig:5}
\end{figure}

In our experiments we used a diverging beam at a large distance, approximating a flat wavefront with the translation of the beam giving the incident angle on the sensing tip of the fibre. This means our wavefront filled the lab, in a beam configuration that was neither relevant nor interesting for astronomy. To validate the applicability of the wavefront sensor to astronomy, we need to estimate the response to incoming light in these systems. To do this we take a model telescope, arranged to retain telecentricty (see  Fig.\,\ref{fig:5}), and subdivide the pupil into subapertures.  
A subaperture $(D_{\mathrm{A}})$ is focussed by the telescope (here represented by a lens of focal length $f_{\mathrm{T}}$). After a further distance $f_{\mathrm{L}}$, a microlens of focal length $f_{\mathrm{L}}$ reimages a pupil subaperture to match the size of the sensing tip of the sensing fibre. This is in many ways similar to how a single element of a Shack-Hartmann WFS is used. The equilateral positioning of the cores in our WFS mean that it will not be sensitive to aberrations that are cylindrically symmetric. Furthermore, the input cores are close enough that higher-order aberrations such as coma and astigmatism will simply act as modifications to the measured tip-tilt of a plane wave.
The magnification of the system, $f_{\mathrm{T}}/f_{\mathrm{L}}$, yields the relation between the angle $\theta_{\mathrm{pupil}}$ on-sky and the angle $\theta_{\mathrm{fibre}}$ detected by the sensor 
\begin{equation}
    \theta_{\mathrm{pupil}}= \frac{n\Lambda}{D_{\mathrm{A}}}\theta_{\mathrm{fibre}}
    \label{eq.4}
\end{equation}
Here, $n$ is the number of sub-samples of the pupil, i.e. the diameter of the subaperture's image on the sensing tip divided by the core spacing, $\Lambda$. Combining this with our previous knowledge of the sensor’s response in different regimes, we can translate this into limits of how and where the wavefront sensor can be used. Using this we can then work out the maximum incoming tilt angle from the pupil for each of these regimes.

\begin{figure}
	\includegraphics[width=\columnwidth]{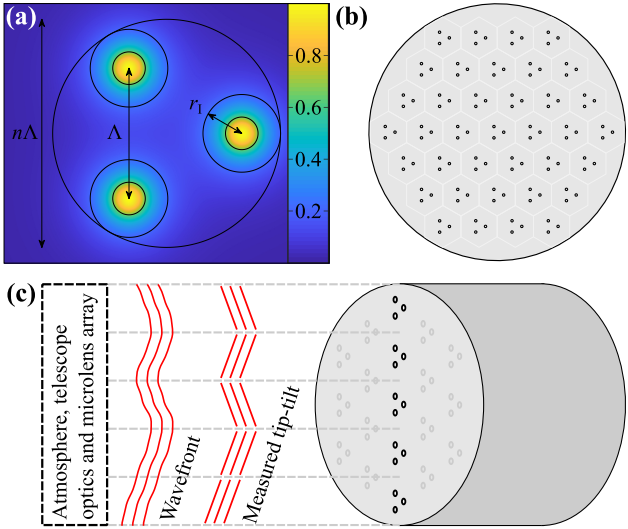}
    \caption{(a) Density plot of the simulated field distribution of the fundamental supermode (the equal superposition of the fields in the three cores), with (black circles in order of increasing size) the core boundaries, the mode radii $r_{\mathrm{I}}$ and a subaperture image of diameter $n\lambda$ that lies one mode radius beyond the centres of the cores. (b) Schematic diagram of a `leopard-print' multicore fibre containing many independent three-core elements, overlaid with the pupil subapertures the cores are sensing. (c) Schematic diagram showing how our proposed fibre facilitates measurements of higher-order wavefront aberrations. An array of three-core sensors measure tip-tilt at points across the wavefront, which can be interpolated to construct a single measure of the higher-order wavefront aberrations. For simplicity, we have depicted this in one dimension, but using all of the nineteen sets of cores would enable a two-dimensional representation of the wavefront to be recovered.}
    \label{fig:6}
\end{figure}

For astronomical wavefront sensing, the throughput of the device is also important. For the case where the subaperture's image covers the tapered fibre's cores and extends one mode radius beyond their centres $(n = 1.75)$, we calculate that 54\,\% of the light through the subaperture is captured by the cores and detected. This coupling efficiency reduces to 40\,\% for a subaperture image that extends two mode-radii beyond the centres of the cores $(n = 2.35)$. These efficiency values were obtained numerically by solving the overlap integral between a plane wave passing through the subaperture and the electric-field distribution of the light confined by the three-core structure at the sensing tip. Figure~\ref{fig:6}\,(a) shows the electric-field distribution at the sensing tip as simulated in the commercial finite-element analysis software, COMSOL Multiphysics. 
The figure shows the modes extending beyond the core diameters when the fibre is tapered, leading to higher coupling efficiencies. For our experimental tapered fibre, the calculated mode field diameter of 14.3\,$\upmu$m is substantial compared to the core separation of 24\,$\upmu$m. It would be possible to increase the coupling efficiency by tapering the sensing tip of the fibre to a smaller diameter. However, the length of the tapered region would need to decrease for the device to maintain the same number of coupling lengths, which at some point would be difficult to achieve.

\subsection{On-sky validation example}
With our laboratory validation of the sensor and theoretical framework, the next step will be to validate the concept on-sky. It would be straightforward to scale the geometry of our WFS design to optimise performance at any wavelength, for example to take advantage of lower noise detectors at visible wavelengths. However, we optimised our prototype for 1550\,nm (the astronomical H-band). Therefore, we consider an AO system capable of providing an IR light path and simultaneously monitoring our performance using its existing infrastructure. We imagine the CANARY AO system on the William Herschel Telescope (WHT) \citep{gendron_2011} as our host AO system. CANARY is an AO demonstrator for the Extremely Large Telescope (ELT), designed with the flexibility to try out different AO flavours and new wavefront sensor types. It has a history of hosting experimental instruments \citep{harris_2015,haffert_2020,sinquin_2020} and crucially is designed to record test data from them in conjunction with AO measurements.

We emulate the CANARY SH-WFS setup, which divides the pupil using a $7\!\times\!7$ microlens array \citep{gendron_2011}, but we perform our wavefront sensing in the IR instead of the visible. Accordingly, our measure of beam coherence, the Fried parameter, $r_{0}$, will be larger. We estimate that $r_{0} =26$\,cm at 1550\,nm by assuming $r_{0}= 15$\,cm at 500\,nm (similar to \citeauthor{gendron_2011}). This means our sensitivity to shorter wavelength aberrations will be lower, limiting our achievable correction when correcting for visible wavelengths.

The WHT has a 4.2-m aperture and a focal ratio of 10.8, giving a focal length of $f_{\mathrm{T}} = 45.36$\,m and CANARY has a pupil subaperture of DA = 60 cm. According to Eq.~\ref{eq.4}, this necessitates a microlens with a focal length of $f_{\mathrm{L}} \sim1.8$\,mm. Since our fibre diameter is not unusually large, this is well within current manufacturing capabilities. In future designs, this microlens could be replaced by 3D printed technologies \citep{dietrich_2017}, allowing faster integration and custom design.

Using the above numbers, and by choosing $n=1.75$, we can show how our sensor would respond if integrated into CANARY. As with our laboratory experiment, we use CANARY's H-band. On-sky, the wavefront sensor fabricated for our experiment would be linear to 51.4\,mas ($\delta = 0.2$), then would be analytically derived until 205.6\,mas ($\delta = 0.8$) and function with the correct sign until 411.3\,mas ($\delta = 1.6$). Accordingly, the sensor is highly sensitive to tilts in the incoming wavefront but with an operating range too small for open-loop operation \citep{glindemann_2000}. As such, we would need the AO loop to be closed with a conventional wavefront sensor before we could begin sensing, though a sufficiently bright star could provide accurate wavefront measurements for small tilts. The sensor can, of course, be made less sensitive by reducing the separation between cores at the sensing tip, $\Lambda$. Making this change would not necessarily diminish the devices' suitability for measuring small angles if the detector setup was still able to distinguish small differences in power between each core. 

The sensor does have a slow wavelength dependence, as shown by the explicit dependence on $\lambda$ in Eq.\,\ref{eq.3}. This can be compensated somewhat by departing from the $z = L_{\mathrm{C}}/4$ length condition within the sinusoid in Eq.\,\ref{eq.3}, since $L_{\mathrm{C}}$ has its own wavelength dependence via the coupling coefficient $C$. In any case the condition of an untilted wavefront $(\theta = 0)$, and hence the sign of the feedback correction needed to attain it, does not depend on wavelength. For the fibre of our experiments, calculations show that the sensitivity of centroid displacement, $\mathbf{r}$, to tilt angle, $\theta$, in Eqs.\,\ref{eq.2} and \ref{eq.3} can be kept greater than 50\,\% of its maximum value over a 350\,nm wavelength range centred on 1550\,nm. In contrast, we expect more strongly-coupled systems like the photonic lantern of \citet{cruz_2021} to necessarily have many more coupling cycles, $z/L_{\mathrm{C}}$, along the tapered section than the $\sim1/4$ of our wavefront sensor, resulting in a strong wavelength dependence and a narrow bandwidth.  

We have considered a single wavefront sensor element measuring tip-tilt across a subaperture of the pupil. To access the full pupil we need an array of several three-core elements. This can be provided monolithically by a multicore fibre containing a `leopard-print' pattern of sets of three cores, shown schematically in Figure~\ref{fig:6}\,(b). The sets of three cores should be far enough apart that light does not couple from one set to its neighbours. The whole fibre can then be tapered to form the array of tip-tilt sensors from which higher-order aberrations can be deduced. We illustrate this concept in  Figure~\ref{fig:6}\,(c), where tip-tilt, as measured by many three-core elements in a single fibre, can be used to reconstruct more complex wavefront aberrations. 

\section{Future Astronomical Use}
The sensor we have detailed here is compact and shows high accuracy over a small operational range and requires only three detector pixels to reconstruct each sensed subaperture. Due to the limited operational range, it would need to be implemented as a closed loop wavefront sensor, sitting behind a correcting DM and performing optimisations on a partially corrected point spread function. The sampling is also set by the Fried parameter, as higher-order aberrations cannot be sensed by the three cores. This makes the sensor dependent on the seeing at the site, with the number of subapertures increasing for larger telescopes.

Due to the small size of the tapered-fibre sensing tip, our all-fibre WFS could be exploited for ELT instrumentation, allowing crowded focal planes to be used simultaneously for both science fibres, integral-field-unit fibres and wavefront sensing fibres without the need for a dichroic to split light between the two. This would allow a reduction in non-common path aberrations caused by vibrations and differences in optics, which is necessary for efficient use of single mode instrumentation.

\section{Conclusions}
We have experimentally demonstrated a tip-tilt wavefront sensor based on a tapered three-core fibre. In this device, a coupled region at the fibre tip region acts as a lens, converting phase information from an incoming wavefront into amplitude information within each of the cores, which can then be routed away from the telescope focal plane. The sensor uses minimal information to infer tip-tilt and is compatible with remote detector arrays. We have also theoretically derived the equations to allow its use in astronomy. 

We have shown that our wavefront sensor is particularly sensitive to small tip and tilt perturbations of the subaperture. Only requiring three cores means that they system can have lower noise than conventional wavefront sensors. In principle, the tilt resolution of our WFS is limited only by the capability of the detector setup to resolve the power difference between each core.

Future work will be conducted on two fronts. First, to theoretically optimize the sensor, with detailed calculations of throughput, signal to noise of the system and analysis of the optimal operating conditions. (Our 3-core wavefront sensor is a proof of concept and has by no means been optimised.) Second, to test a prototype in a telescope simulator and on sky. This will allow us to confirm the laboratory performance in realistic conditions.

\section*{Acknowledgements}

This work was supported by the European Union’s Horizon 2020 research and innovation program under grant no. 730890 (OPTICON – Optical Infrared Coordination Network for Astronomy).

%%%%%%%%%%%%%%%%%%%%%%%%%%%%%%%%%%%%%%%%%%%%%%%%%%
\section*{Data Availability}

Data underlying the results presented are available at \citep{wright_2022}.



\bibliographystyle{mnras}
\bibliography{bibliography} % if your bibtex file is called example.bib

\begin{thebibliography}{}
\makeatletter
\relax
\def\mn@urlcharsother{\let\do\@makeother \do\$\do\&\do\#\do\^\do\_\do\%\do\~}
\def\mn@doi{\begingroup\mn@urlcharsother \@ifnextchar [ {\mn@doi@}
  {\mn@doi@[]}}
\def\mn@doi@[#1]#2{\def\@tempa{#1}\ifx\@tempa\@empty \href
  {http://dx.doi.org/#2} {doi:#2}\else \href {http://dx.doi.org/#2} {#1}\fi
  \endgroup}
\def\mn@eprint#1#2{\mn@eprint@#1:#2::\@nil}
\def\mn@eprint@arXiv#1{\href {http://arxiv.org/abs/#1} {{\tt arXiv:#1}}}
\def\mn@eprint@dblp#1{\href {http://dblp.uni-trier.de/rec/bibtex/#1.xml}
  {dblp:#1}}
\def\mn@eprint@#1:#2:#3:#4\@nil{\def\@tempa {#1}\def\@tempb {#2}\def\@tempc
  {#3}\ifx \@tempc \@empty \let \@tempc \@tempb \let \@tempb \@tempa \fi \ifx
  \@tempb \@empty \def\@tempb {arXiv}\fi \@ifundefined
  {mn@eprint@\@tempb}{\@tempb:\@tempc}{\expandafter \expandafter \csname
  mn@eprint@\@tempb\endcsname \expandafter{\@tempc}}}

\bibitem[\protect\citeauthoryear{Akyol, Hagag, Sivaprasad  \& Lotery}{Akyol
  et~al.}{2021}]{akyol_2021}
Akyol E.,  Hagag A.~M.,  Sivaprasad S.,   Lotery A.~J.,  2021, \mn@doi [Eye]
  {10.1038/s41433-020-01286-z}, 35, 244

\bibitem[\protect\citeauthoryear{Birks, Gris-S{\'a}nchez, Yerolatsitis,
  Leon-Saval  \& Thomson}{Birks et~al.}{2015}]{birks_2015}
Birks T.~A.,  Gris-S{\'a}nchez I.,  Yerolatsitis S.,  Leon-Saval S.,   Thomson
  R.~R.,  2015, \mn@doi [Adv. Opt. Photonics] {10.1364/AOP.7.000107}, 7, 107

\bibitem[\protect\citeauthoryear{Booth}{Booth}{2014}]{booth_2014}
Booth M.~J.,  2014, \mn@doi [Light Sci. Appl.] {10.1038/lsa.2014.46}, 3, e165

\bibitem[\protect\citeauthoryear{{Chauvin, G.}, {Lagrange, A.-M.}, {Dumas, C.},
  {Zuckerman, B.}, {Mouillet, D.}, {Song, I.}, {Beuzit, J.-L.}  \& {Lowrance,
  P.}}{{Chauvin, G.} et~al.}{2004}]{chauvin_2004}
{Chauvin, G.} {Lagrange, A.-M.} {Dumas, C.} {Zuckerman, B.} {Mouillet, D.}
  {Song, I.} {Beuzit, J.-L.}  {Lowrance, P.} 2004, \mn@doi [A\&A]
  {10.1051/0004-6361:200400056}, 425, L29

\bibitem[\protect\citeauthoryear{Corrigan, Harris, Thomson, MacLachlan,
  Allington-Smith, Myers  \& Morris}{Corrigan et~al.}{2016}]{corrigan_2016}
Corrigan M.,  Harris R.~J.,  Thomson R.~R.,  MacLachlan D.~G.,  Allington-Smith
  J.,  Myers R.,   Morris T.,  2016, in Marchetti E.,  Close L.~M.,   Véran
  J.-P.,  eds,  Proc. SPIE Conf. Ser. Vol. 9909, Adaptive Optics Systems V.
  SPIE, pp 1848 -- 1855, \mn@doi{10.1117/12.2230568}

\bibitem[\protect\citeauthoryear{Corrigan, Morris, Harris  \& Anagnos}{Corrigan
  et~al.}{2018}]{corrigan_2018}
Corrigan M.~K.,  Morris T.~J.,  Harris R.~J.,   Anagnos T.,  2018, in Close
  L.~M.,  Schreiber L.,   Schmidt D.,  eds,  Proc. SPIE Conf. Ser. Vol. 10703,
  Adaptive Optics Systems VI. SPIE, pp 1313 -- 1320,
  \mn@doi{10.1117/12.2311336}

\bibitem[\protect\citeauthoryear{Cruz-Delgado et~al.,}{Cruz-Delgado
  et~al.}{2021}]{cruz_2021}
Cruz-Delgado D.,  et~al., 2021, \mn@doi [Opt, Lett.] {10.1364/OL.430761}, 46,
  3292

\bibitem[\protect\citeauthoryear{Daigle, Kamali, Ch{\^a}teauneuf, Tremblay,
  Th{\'e}berge, Dubois, Roy  \& Chin}{Daigle et~al.}{2009}]{daigle_2009}
Daigle J.-F.,  Kamali Y.,  Ch{\^a}teauneuf M.,  Tremblay G.,  Th{\'e}berge F.,
  Dubois J.,  Roy G.,   Chin S.,  2009, \mn@doi [Appl. Phys. B]
  {10.1007/s00340-009-3713-7}, 97, 701

\bibitem[\protect\citeauthoryear{Davies \& Kasper}{Davies \&
  Kasper}{2012}]{davies_2012}
Davies R.,  Kasper M.,  2012, \mn@doi [Annu. Rev. Astron. Astr.]
  {10.1146/annurev-astro-081811-125447}, 50, 305

\bibitem[\protect\citeauthoryear{Dietrich, Harris, Blaicher, Corrigan, Morris,
  Freude, Quirrenbach  \& Koos}{Dietrich et~al.}{2017}]{dietrich_2017}
Dietrich P.-I.,  Harris R.~J.,  Blaicher M.,  Corrigan M.~K.,  Morris T.~J.,
  Freude W.,  Quirrenbach A.,   Koos C.,  2017, \mn@doi [Opt. Express]
  {10.1364/OE.25.018288}, 25, 18288

\bibitem[\protect\citeauthoryear{Gendron et~al.,}{Gendron
  et~al.}{2011}]{gendron_2011}
Gendron E.,  et~al., 2011, \mn@doi [A\&A] {10.1051/0004-6361/201116658}, 529,
  L2

\bibitem[\protect\citeauthoryear{Ghez, Klein, Morris  \& Becklin}{Ghez
  et~al.}{1998}]{ghez_1998}
Ghez A.~M.,  Klein B.~L.,  Morris M.,   Becklin E.~E.,  1998, \mn@doi [ApJ]
  {10.1086/306528}, 509, 678

\bibitem[\protect\citeauthoryear{Glindemann, Hippler, Berkefeld  \&
  Hackenberg}{Glindemann et~al.}{2000}]{glindemann_2000}
Glindemann A.,  Hippler S.,  Berkefeld T.,   Hackenberg W.,  2000, \mn@doi
  [Exp. Astron.] {10.1023/A:1008116831367}, 10, 5

\bibitem[\protect\citeauthoryear{Gonsalves \& Chidlaw}{Gonsalves \&
  Chidlaw}{1979}]{gonsalves_1979}
Gonsalves R.~A.,  Chidlaw R.,  1979, in Tescher A.~G.,  ed.,  Proc. SPIE Conf.
  Ser. Vol. 0207, Applications of Digital Image Processing III. SPIE, pp 32 --
  39, \mn@doi{10.1117/12.958223}

\bibitem[\protect\citeauthoryear{Gregory, Heine, K{\"a}mpfner, Lange, Lutzer
  \& Meyer}{Gregory et~al.}{2012}]{gregory_2012}
Gregory M.,  Heine F.~F.,  K{\"a}mpfner H.,  Lange R.,  Lutzer M.,   Meyer R.,
  2012, \mn@doi [Opt. Eng.] {10.1117/1.OE.51.3.031202}, 51, 031202

\bibitem[\protect\citeauthoryear{Guyon}{Guyon}{2018}]{guyon_2018}
Guyon O.,  2018, \mn@doi [Annu. Rev. Astron. Astr.]
  {10.1146/annurev-astro-081817-052000}, 56, 315

\bibitem[\protect\citeauthoryear{Haffert et~al.,}{Haffert
  et~al.}{2020}]{haffert_2020}
Haffert S.~Y.,  et~al., 2020, \mn@doi [J. Astron. Telesc. Instrum. Syst.]
  {10.1117/1.JATIS.6.4.045007}, 6, 1

\bibitem[\protect\citeauthoryear{Harris et~al.,}{Harris
  et~al.}{2015}]{harris_2015}
Harris R.~J.,  et~al., 2015, \mn@doi [MNRAS] {10.1093/mnras/stv410}, 450, 428

\bibitem[\protect\citeauthoryear{Norris, Wei, Betters, Wong  \&
  Leon-Saval}{Norris et~al.}{2020}]{norris_2020}
Norris B.~R.,  Wei J.,  Betters C.~H.,  Wong A.,   Leon-Saval S.~G.,  2020,
  \mn@doi [Nat. Commun.] {10.1038/s41467-020-19117-w}, 11, 1

\bibitem[\protect\citeauthoryear{Paxman, Thelen, Murphy, Gleichman  \&
  III}{Paxman et~al.}{2007}]{paxman_2007}
Paxman R.~G.,  Thelen B.~J.,  Murphy R.~J.,  Gleichman K.~W.,   III J. A.~G.,
  2007, in Carreras R.~A.,  Gonglewski J.~D.,   Rhoadarmer T.~A.,  eds,  Proc.
  SPIE Conf. Ser. Vol. 6711, Advanced Wavefront Control: Methods, Devices, and
  Applications V. SPIE, pp 8 -- 22, \mn@doi{10.1117/12.734665}

\bibitem[\protect\citeauthoryear{Platt \& Shack}{Platt \&
  Shack}{2001}]{platt_2001}
Platt B.~C.,  Shack R.,  2001, \mn@doi [J. Refract. Surg.]
  {10.3928/1081-597X-20010901-13}, 17, S573

\bibitem[\protect\citeauthoryear{Ragazzoni}{Ragazzoni}{1996}]{ragazzoni_1996}
Ragazzoni R.,  1996, \mn@doi [J. Mod. Optic.] {10.1080/09500349608232742}, 43,
  289

\bibitem[\protect\citeauthoryear{Rigaut}{Rigaut}{2015}]{rigaut_2015}
Rigaut F.,  2015, \mn@doi [Publ. Astron. Soc. Pac.] {10.1086/684512}, 127, 1197

\bibitem[\protect\citeauthoryear{Roddier}{Roddier}{1988}]{roddier_1988}
Roddier F.,  1988, \mn@doi [Appl. Opt.] {10.1364/AO.27.001223}, 27, 1223

\bibitem[\protect\citeauthoryear{Russell}{Russell}{2003}]{russell_2003}
Russell P.,  2003, \mn@doi [Science] {10.1126/science.1079280}, 299, 358

\bibitem[\protect\citeauthoryear{Sauvage, Fusco, Rousset  \& Petit}{Sauvage
  et~al.}{2007}]{sauvage_2007}
Sauvage J.-F.,  Fusco T.,  Rousset G.,   Petit C.,  2007, \mn@doi [JOSA A]
  {10.1364/JOSAA.24.002334}, 24, 2334

\bibitem[\protect\citeauthoryear{Sinquin et~al.,}{Sinquin
  et~al.}{2020}]{sinquin_2020}
Sinquin B.,  et~al., 2020, \mn@doi [MNRAS] {10.1093/mnras/staa2562}, 498, 3228

\bibitem[\protect\citeauthoryear{Tian \& Huhns}{Tian \&
  Huhns}{1986}]{tian_1986}
Tian Q.,  Huhns M.~N.,  1986, \mn@doi [Comput. graph. image process.]
  {10.1016/0734-189X(86)90028-9}, 35, 220

\bibitem[\protect\citeauthoryear{Valente, Rativa  \& Vohnsen}{Valente
  et~al.}{2015}]{Valente_15}
Valente D.,  Rativa D.,   Vohnsen B.,  2015, \mn@doi [Opt. Express]
  {10.1364/OE.23.013005}, 23, 13005

\bibitem[\protect\citeauthoryear{Weyrauch \& Vorontsov}{Weyrauch \&
  Vorontsov}{2005}]{weyrauch_2005}
Weyrauch T.,  Vorontsov M.~A.,  2005, \mn@doi [Appl. Opt.]
  {10.1364/AO.44.006388}, 44, 6388

\bibitem[\protect\citeauthoryear{Wright, Yerolatsitis, Kerrianne, Harris  \&
  Birks}{Wright et~al.}{2022}]{wright_2022}
Wright T.~A.,  Yerolatsitis S.,  Kerrianne H.,  Harris R.~J.,   Birks T.~A.,
  2022, Dataset for "All-fibre wavefront sensor", \mn@doi{10.15125/BATH-01135}

\bibitem[\protect\citeauthoryear{Zawadzki et~al.,}{Zawadzki
  et~al.}{2005}]{zawadzki_2005}
Zawadzki R.~J.,  et~al., 2005, \mn@doi [Opt. Express] {10.1364/OPEX.13.008532},
  13, 8532

\makeatother
\end{thebibliography}

%\begin{thebibliography}{99}
%\end{thebibliography}

%%%%%%%%%%%%%%%%%%%%%%%%%%%%%%%%%%%%%%%%%%%%%%%%%%

%%%%%%%%%%%%%%%%% APPENDICES %%%%%%%%%%%%%%%%%%%%%

\appendix

\section{Analytical relationship between wavefront tip-tilt and centroid position}

The behaviour of the coupling section at the input of the fibre wavefront sensor can be analysed using its supermodes. These are the combinations of amplitudes and phases in (the fundamental modes of) each core which propagate along the fibre without change except for an overall phase $\sim e^{i\beta z} $, where $\beta$ is the supermode's propagation constant. For three identical weakly-coupled equally-spaced cores, the state vectors describing the amplitudes and phases in the three cores for each supermode A, B and C can be written 
\begin{equation}
\mathbf{a}_{\mathrm{A}}=\frac{1}{\sqrt{3}}\begin{pmatrix}
1 \\
1 \\
1
\end{pmatrix} \quad \mathbf{a}_{\mathrm{B}}=\frac{1}{\sqrt{6}}\begin{pmatrix}
2 \\
-1 \\
-1
\end{pmatrix} \quad \mathbf{a}_{\mathrm{C}}=\frac{1}{\sqrt{2}}\begin{pmatrix}
0 \\
1 \\
-1
\end{pmatrix},
    \label{eq.a1}
\end{equation}
with the corresponding propagation constants
\begin{align}
\beta_{\mathrm{A}} = \beta_0 + 2C, \label{eq.a2} 
\end{align}
and
\begin{equation}
    \beta_{\mathrm{B}}=\beta_{\mathrm{C}}=\beta_{0}-C,
    \label{eq.a3}
\end{equation}
where $\beta_0$ is the propagation constant of one core in isolation, and $C$ is the coupling coefficient between a pair of cores.

The input field is written as a superposition of the three supermodes, which are allowed to propagate through distance $z$ to the end of the coupling region and thereby accumulate different phases according to their $\beta$ values. The square modulus of the output amplitude in each core gives the normalised power $P$ that emerges from that core.

To deduce the coupling characteristic in Fig.\,\ref{fig:2} that were used to ensure that the coupling region was the right length, we take the input state vector to be $\mathbf{a} = (1, 0, 0)$, i.e. light in core 1 only. The output powers are then
\begin{align}
    P_1&=\left[5+4\cos(3Cz)\right]/9,\label{eq.a4} \\ P_2&=P_3=\left[2-2\cos(3Cz)\right]/9.\label{eq.a5}
\end{align}
The coupling length $L_{\mathrm{C}}$ is the distance, $z$, where the powers return to the input condition of $P_1$ = 1. It is given by $3CL_{\mathrm{C}}$ = $2\pi$, and can be determined readily from the measured output powers using Eq.~(\ref{eq.a4}--\ref{eq.a5}). $P_1$ is minimised when $3Cz = \pi$, corresponding to $z = L_{\mathrm{C}}/2$.

In a non-uniform structure, such as a real fibre taper with a transition along which coupling gradually falls to zero, $C$ is position-dependent and $L_{\mathrm{C}}$ becomes ill-defined. However, the expressions in this paper remain correct if $Cz$ is generalised to the integral of $C\mathrm{d}z$, with the ratio $z/L_{\mathrm{C}} = 3Cz/2\pi$ being well-defined as the overall number of complete coupling cycles along the whole tapered section.

To deduce the response of the structure to a tilted wavefront, we use the transverse wavevector $\mathbf{k}_{\mathrm{T}} = k_0\theta(\cos\alpha, \sin\alpha)$ in the plane of the fibre endface with the geometry of Fig.\,\ref{fig:1}\,(a). Here, $k_0$ is the free-space wavenumber, $k_0=2\pi/\lambda$. We assume small $\theta$, and evaluate the phases $\mathbf{k}_{\mathrm{T}}\cdot \mathbf{r}$ at the centres of the cores to determine the input state vector
\begin{equation}
    \mathbf{a} = \begin{pmatrix}
e^{i\delta_0\cos \alpha} \\
e^{i\delta_0\cos(\alpha-2\pi/3)} \\
e^{i\delta_0\cos(\alpha-2\pi/3)}
\end{pmatrix},
\label{eq.a6}
\end{equation}
where
\begin{equation}
    \delta_0 = \frac{k_0\theta\Lambda}{\sqrt{3}}.
    \label{eq.a7}
\end{equation}

After some effort, the supermode analysis yields an analytic expression for the output power in core 1:
\begin{equation}
\begin{aligned}
    P_{1}= 1+\frac{2}{9}\left[1-\cos \left(\frac{2 \pi z}{L_{C}}\right)\right]\big\{2 \cos \left[\delta_{0}\left(\chi_{-}-\chi_{+}\right)\right]-\\
    \cos \left[\delta_{0}\left(\chi_{0}-\chi_{-}\right)\right]-\cos \left[\delta_{0}\left(\chi_{+}-\chi_{0}\right)\right]\big\} \\
+\frac{2}{3} \sin \left(\frac{2 \pi z}{L_{C}}\right)\big\{\sin \left[\delta_{0}\left(\chi_{0}-\chi_{-}\right)\right]-\sin \left[\delta_{0}\left(\chi_{+}-\chi_{0}\right)\right]\big\},
    \label{eq.a8}
\end{aligned}
\end{equation}
where
\begin{align}
    \chi_0&=\cos\alpha, \label{eq.a9} \\  
    \chi_\pm&= \cos (\alpha\pm2\pi/3).\label{eq.a10}
\end{align}
The powers in cores 2 and 3 are given in turn by the cyclic rotation $\chi_0\!\rightarrow\!\chi_-\!\rightarrow\!\chi_+\!\rightarrow\!\chi_0$, which is equivalent to subtracting $2\pi/3$ from each occurrence of $\alpha$. These powers can be expressed using the centroid of Eq.~(\ref{eq.1}) in polar coordinates as
\begin{align}
    \rho&=\frac{\sqrt{\left(P_{1}-P_{2}\right)^{2}+\left(P_{2}-P_{3}\right)^{2}+\left(P_{3}-P_{1}\right)^{2}}}{\sqrt{2}\, \left(P_{1}+P_{2}+P_{3}\right)},\label{eq.a11} \\
    \cos \phi&=\frac{\left(P_{1}-P_{2}\right)+\left(P_{1}-P_{3}\right)}{\sqrt{2}\,  \sqrt{\left(P_{1}-P_{2}\right)^{2}+\left(P_{2}-P_{3}\right)^{2}+\left(P_{3}-P_{1}\right)^{2}}},\label{eq.a12} \\
    \sin \phi&=\frac{\sqrt{3} \,\left(P_{1}-P_{3}\right)}{\sqrt{2}\, \sqrt{\left(P_{1}-P_{2}\right)^{2}+\left(P_{2}-P_{3}\right)^{2}+\left(P_{3}-P_{1}\right)^{2}}}.\label{eq.a13}
\end{align}

Calculation of $\phi$ is simpler via $\tan\phi$, although the sign of $\cos\phi$ or $\sin\phi$ is needed to get $\phi$ in the right quadrant.

To first order in $\delta_{0}$,  Eq.~(\ref{eq.a8}) becomes
\begin{align}
    P_{1}&=1+2 \delta_{0} \cos \alpha \sin \left(2 \pi z / L_{C}\right)\label{eq.a14}, \\
    P_{2}&=1+2 \delta_{0} \cos (\alpha-2 \pi / 3) \sin \left(2 \pi z / L_{C}\right)\label{eq.a15}, \\
    P_{3}&=1+2 \delta_{0} \cos (\alpha+2 \pi / 3) \sin \left(2 \pi z / L_{C}\right)\label{eq.a16},
    \end{align}
and application of Eq.~(\ref{eq.a11}--\ref{eq.a13}) yields Eq.~(\ref{eq.2}) and Eq.~(\ref{eq.3}).

Expressions of similar form can be derived for cases where the coupling between the cores is not weak. For example, in a typical photonic lantern, the fibre is tapered small enough that the fibre cores lose their identities and the tapered section acts like a single multi-mode fibre core \citep{birks_2015}. In such cases, $C$ is not defined but is replaced in the analysis by $(\beta_{01} - \beta_{11})/3$, where $\beta_{lm}$ is the propagation constant of the $\mathrm{LP}_{lm}$ mode. This is typically much greater than $C$ for a pair of well-separated cores, meaning that the multimode section of any realistic photonic lantern will be considerably longer than its $L_{\mathrm{C}}$. This makes it difficult to meet the $z = L_{\mathrm{C}}/4$ condition, and produces a strongly `beating' wavelength dependence. For this reason, we do not consider this case further.
%%%%%%%%%%%%%%%%%%%%%%%%%%%%%%%%%%%%%%%%%%%%%%%%%%

% Don't change these lines
\bsp	% typesetting comment
\label{lastpage}
\end{document}